\pdfoutput=1
\documentclass[reprint, aps, prd, preprintnumbers, amsmath, amssymb, superscriptaddress, nofootinbib]{revtex4-1}
\usepackage{amssymb,amsmath,bbold,slashed}
\usepackage{graphicx,soul}
\usepackage[usenames,dvipsnames]{xcolor}
\usepackage[unicode=true,pdfusetitle,
 bookmarks=true,bookmarksnumbered=false,bookmarksopen=false,citecolor=Turquoise,
 breaklinks=false,pdfborder={0 0 1},backref=false,colorlinks=true,pdfpagemode=FullScreen]
 {hyperref}

\begin{document}

\title{Effect of Ultralight Dark Matter on $g-2$ of the Electron}

\author{Jason L. Evans}
\email[]{jlevans@sjtu.edu.cn}
\affiliation{T. D. Lee Institute and School of Physics and Astronomy, Shanghai Jiao Tong University, Shanghai 200240, China}

\begin{abstract}
If dark matter is ultralight, the number density of dark matter is very high and the techniques of zero-temperature field theory are no longer valid. The dark matter number density modifies the vacuum giving it a non-negligible particle occupation number. For fermionic dark matter, this occupation number can be no larger than one.  However, in the case of bosons the occupation number is unbounded. If there is a large occupation number, the Bose enhancement needs to be taken into consideration for any process involving particles which interact with the dark matter. Because the occupation number scales inversely with the dark matter mass, this effect is most prominent for ultralight dark matter. In fact, the Bose enhancement effect from the background is so significant for ultralight dark matter that, if dark matter is a dark photon, the correction to the anomalous magnetic moment is larger than experimental uncertainties for a mixing parameter of order $10^{-16}$ and a dark photon mass of order $10^{-20}$ eV. Furthermore, the constraint on the mixing parameter scales linearly with the dark photon mass and so new significant constraints can be placed on the dark matter mass all the way up to $10^{-14}$ eV. Future experiments measuring $g-2$ will probe even smaller gauge mixing parameters.  
\end{abstract}

\maketitle

\section{Introduction}
In quantum field theory, we often assume that the process we are considering takes place in empty space. Under this assumption, the lowering and raising operators of the quantum field lead to the following relation
\begin{eqnarray}
\langle 0|a(k)a^\dagger (k')|0\rangle =(2\pi)^3(2k_0)\delta^3(k-k')~,
\end{eqnarray}
when bosonic operators $a(k)$ are considered. This relation can then be used to calculate the, well known, propagator for a boson:
\begin{eqnarray}
G(k)=\frac{i}{k^2-m^2+i\epsilon}~.
\end{eqnarray}

However, if the process of interest is happening on a background, we are no longer considering the propagation from vacuum to vacuum. We are instead considering propagation from an $|\bar n(k)\rangle$ boson state to another $|\bar n(k)\rangle$ boson state. In this case, the relevant expectation value is
\begin{eqnarray}
\langle \bar n|a(k)a^\dagger (k')|\bar n\rangle =(2\pi)^3(2k_0)(1+\bar n(k))\delta^3(k-k')~,
\end{eqnarray}
which leads to a propagator of
\begin{eqnarray}
G(k)=\frac{i}{k^2-m^2+i\epsilon} -2\pi \bar n(k)\delta(k^2-m^2)~.\label{eq:Gkn}
\end{eqnarray}
This is exactly what happens in finite temperature field theory. For example, if we were to take $\bar n(k)$ to be the Bose-Einstein distribution of some plasma, like we often do when we consider the early universe, this would merely be the standard propagator of finite temperature field theory in the real-time formulation. However, in the above discussion, we made no mention of the properties of $\bar n(k)$. We only required that it represented the number of background fields present, with a given momentum. This means, we are free to consider quantum processes for other backgrounds described by other $\bar n(k)$. In this letter, I will consider $\bar n(k)$ to be the profile of some bosonic background dark matter. If there are many dark matter particles present, which is generally the case when dark matter is ultralight, the background contribution to the propagator will have a non-trivial effect on all loop processes involving the dark matter. 

Processes like that considered here are often only marginally enhanced due to a cancellation of the Bose enhanced pieces\footnote{For an example where complete cancellation of the Bose enhancement occurs, see \cite{Donoghue:1983qx}.}. This is due to the fact that these pieces are often associated with infrared divergences. In fact, in this calculation, the leading order contribution also cancel since it is associated with an infrared divergence. However, the effect is so larger that the subleading contribution is still quite significant.   


In this letter, I will present a proof of concept. I will consider corrections to the anomalous magnetic moment\footnote{The muon $g-2$ measurement also has been used to constraint bosonic dark matter in other scenario \cite{Janish:2020knz,Graham:2020kai}.} of the electron from a dark photon dark matter background. This will lead to new very strong constraints for dark matter masses less than about $10^{-14}$ eV. 

\section{The Vertex Correction}
In this section, I will calculate the anomalous magnetic moment of the electron in a background of dark photons\footnote{For some of the original work on dark photons see \cite{Holdom:1985ag,Fayet:1990wx,Fayet:1980ad,Fayet:1980rr,Okun:1982xi,Georgi:1983sy}. For reviews on the subject and constraints see \cite{Raggi:2015yfk,Deliyergiyev:2015oxa,Alekhin:2015byh,Alexander:2016aln,Beacham:2019nyx,Proceedings:2012ulb,Essig:2013lka,Caputo:2021eaa}.}. A very similar calculation was done in \cite{Donoghue:1984zz,Fujimoto:1982np,Peressutti:1981jg} for a Bose distribution of standard model photons. I will borrow much of the techniques for this calculation from \cite{Donoghue:1984zz}. Their techniques can be applied here, since they rely on the form of the propagator, which is effectively the same, and not on the details of $\bar n(k)$.


In this calculation, I take the propagator of the dark photon to be
\begin{eqnarray}
D_{\mu\nu}(k)=-g_{\mu\nu}\left[\frac{i}{p^2-m^2+i\epsilon}+2\pi\bar n(k)\delta(p^2-m^2)\right] ~,
\end{eqnarray} 
I will do this calculation in the Feynman-'t Hooft gauge. Because only the Z boson is associated with the neutral Goldstone boson of the standard model, the mixing of the dark U(1) Goldstone and the SM neutral Goldstone boson will be the same as the mixing of the Z and dark photon. This mixing is of order $m_{DM}^2/M_Z^2$ \cite{Fukuda:1974kn}, and the dark Goldstone boson's interaction with the electron can be safely ignored.

For now, $\bar n(k)$ remains unspecified.  This will permit me to hide my ignorance of the dark matter profile in $\bar n(k)$. At the end of my calculation, I will give an estimate of $\bar n(k)$. 

\begin{figure}[!t]
\centering
\includegraphics[width=3.5cm,trim={0in .4in 0in .4in},clip]{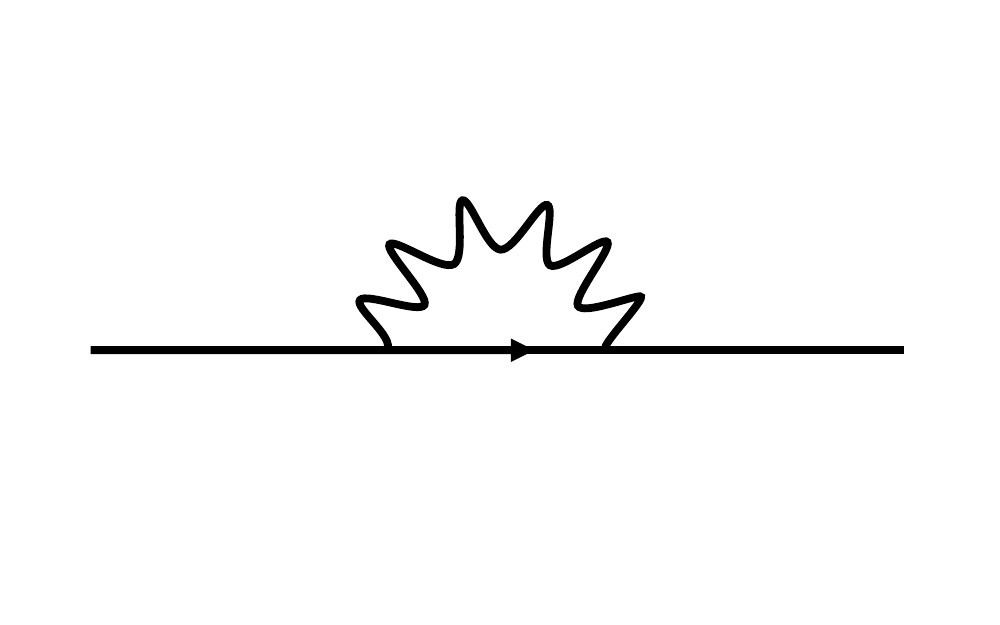}
\includegraphics[width=3cm]{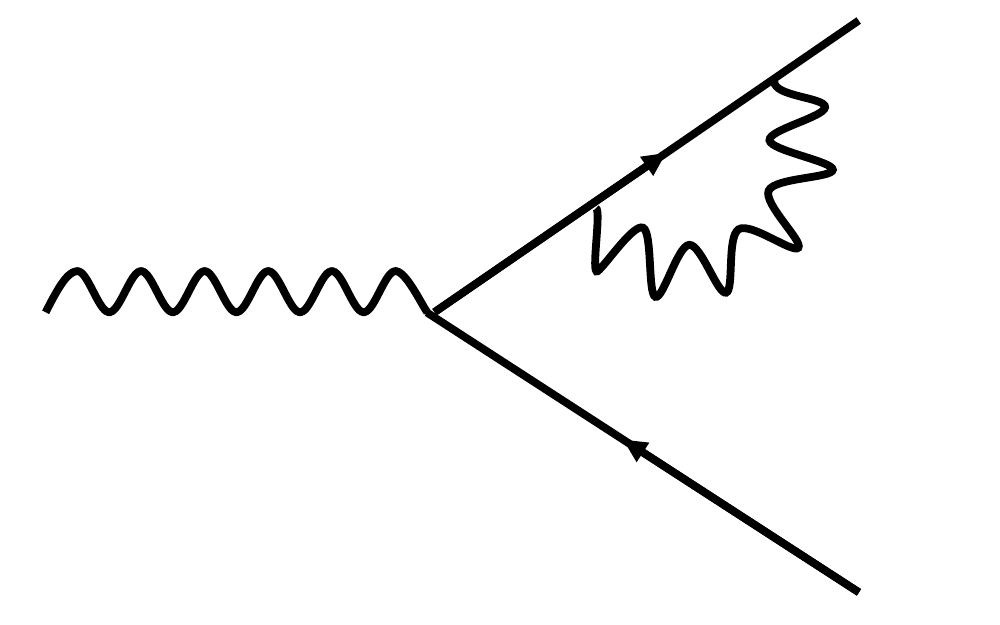}\\
\vspace{0.25in}
\includegraphics[width=3cm]{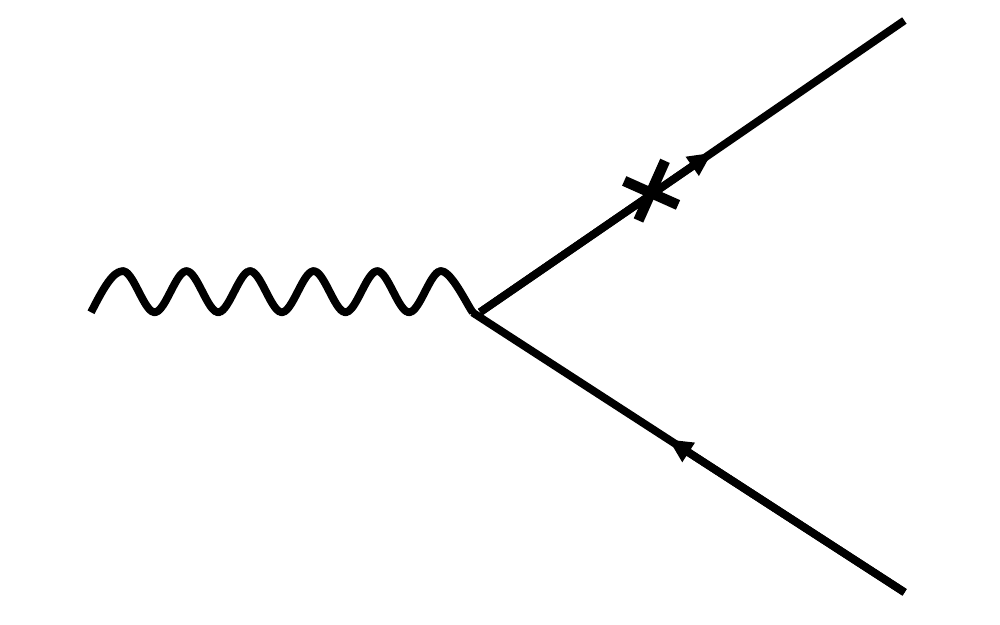}
\includegraphics[width=3cm]{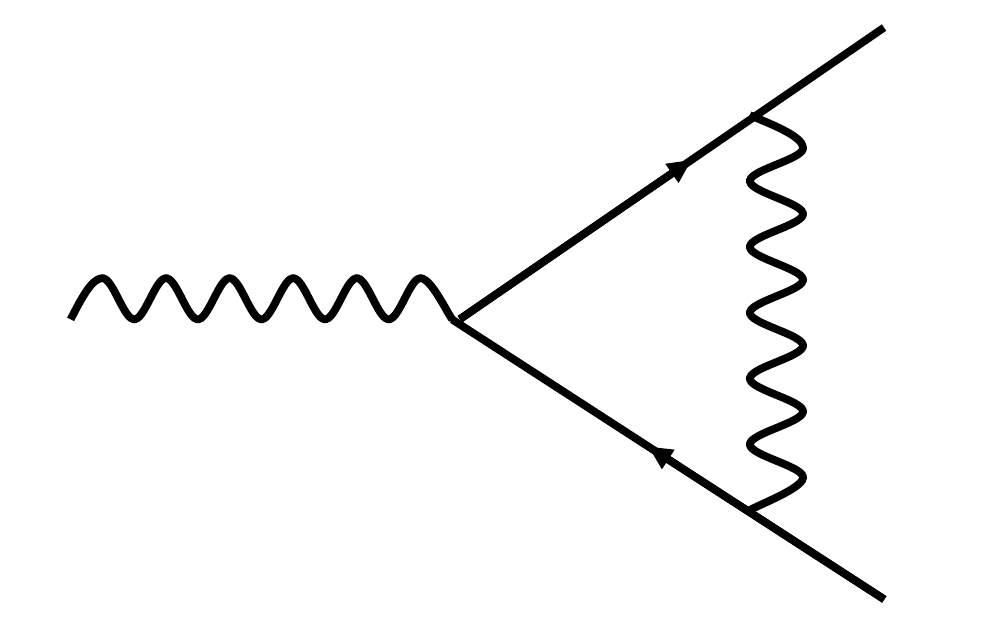}
\caption{\it The relevant Feynman diagrams.}
\label{fig:VerCorrections}
\end{figure}

Now, I calculate the background dependent self-energy correction. This amounts to calculating the top left diagram shown in Fig. (\ref{fig:VerCorrections}). I will only keep the part of the propagator with $\bar n(k)$, since the contribution of the other piece is well known. I separate the self-energy contribution into three unique pieces,
\begin{eqnarray}
\Sigma_n(k)=B(k)+C(k)\left(\slashed{k}-m_e\right)+\slashed{D}(k)~,\label{eq:waveFuncLoop}
\end{eqnarray}
where 
\begin{eqnarray}
B(k)=~~2 e^2\chi^2\int \frac{d^4\Pi_q}{(2\pi)^3} \frac{m_e}{q^2+2q\cdot k +k^2-m_e^2} ~, \label{eq:Bofk}\\
C(k)= -2 e^2\chi^2\int \frac{d^4\Pi_q}{(2\pi)^3} \frac{1}{q^2+2q\cdot k +k^2-m_e^2}~,\label{eq:Cofk}\\
D^\mu(k)= -2 e^2\chi^2\int \frac{d^4\Pi_q}{(2\pi)^3} \frac{q_\mu}{q^2+2q\cdot k +k^2-m_e^2}~, \label{eq:Dofk}
\end{eqnarray}
and
\begin{eqnarray}
d^4\Pi_q= d^4q~\bar n(E_q)\delta(q^2-m_e^2)~.
\end{eqnarray}
$B(k)$ and $D^{\mu}(k)$ lead to a mass correction of the electron, $\delta m_e=B(k)+k_\mu D^{\mu}(k)/m_e$. $C(k)$ corresponds to some sort of wave-function renormalization, which must cancel. The $\chi$ in the expression for $B(k)$, $C(k)$ and $D^\mu(k)$ is the gauge kinetic mixing parameter.

Next, I calculate the three other diagrams in Fig. (\ref{fig:VerCorrections}). These contributions are added to the tree level contribution times the wave function renormalization 
\begin{eqnarray}
Z_2^{-1}\!= 1+C(k)+\frac{m_e}{E}\frac{d}{dE}\!\left[B(k)+\frac{k_\mu D^\mu(k)}{m_e}\right]\!-\frac{D^0(k)}{E} ~.
\end{eqnarray}
to get a total contribution\footnote{For more details, please see Appendix \ref{App:Loop}. For a similar treatment of $Z_2^{-1}$ see \cite{Donoghue:1984zz}.}   
\begin{eqnarray}
&&iM_{TOT\mu}= -ie \bar u_n(\bar k)\left[ \gamma_\mu\left[1\frac{}{} \right. \right.\label{eq:MTOT}\\ 
&& \nonumber  -\frac{1}{2}\frac{1}{E}\frac{d}{dE}\left(m_eB(k)+k_\nu D^\nu(k)\right)  \left.
 +\frac{1}{2}\frac{D^0(k)}{E}+(k\leftrightarrow \bar k)
\right]\\
&& \nonumber +\left[\frac{1}{2}\frac{d }{dk_\mu}\left[B(k)+ \frac{k_\nu D^\nu(k)}{m_e}\right] -\frac{D^\mu(k)}{2m_e} \right. \\  \nonumber 
&& \left.\left. + \frac{\left[\gamma_\alpha,\gamma_\nu\right]_-\Delta k_\alpha}{8m_e}\frac{d D^\nu(k)}{dk_\mu}+\left(k\leftrightarrow \bar k\right)\right]+F_\mu(\Delta k)\right]u_n(k)~,
\end{eqnarray}
where $k\leftrightarrow \bar k$ above leaves $\Delta k$ in the fourth line unchanged and I have neglected terms of higher order than $\Delta k$ and $m_{DM}^2$. In the above expression, after derivatives are performed, I take $k^2=m_e^2$. $F_\mu(\Delta k)$ is defined as  
\begin{eqnarray}
\!\!\!\!\! F_\mu(\Delta k)\quad\quad\quad\quad\quad\quad\quad\quad\quad\quad\quad\quad\quad\quad\quad\quad\quad\quad\quad \\ \nonumber 
=  -e^2\chi^2\int \frac{d^4\Pi_q}{(2\pi)^3}\frac{2\left(\slashed{q}+m_e\right)\left[\Delta \slashed{k},\gamma_\mu\right]-4\gamma_\mu\left[\Delta k_\alpha q^\alpha\right]}{\left(q^2+2 q_\nu k^\nu\right)^2} ~, \label{eq:FDelK}
\end{eqnarray}
Importantly, the wave function renormalization has canceled from Eq. (\ref{eq:MTOT}). After simplification, all pieces associated with the IR divergence of the massless dark photon limit will cancel as well. In fact, if I take the dark photon to be massless, $\chi=1$, and $\bar n(k)$ to be a Bose-Einstein distribution, my calculation of the diagrams in Fig. (\ref{fig:VerCorrections}) exactly reproduces that found in \cite{Donoghue:1984zz}.

\section{Charge Non-Renormalization and Gauge Invariances}
To validate my treatment of the background, I show that the charge remains unrenormalized \cite{Yee:1984wt} and the Ward identities are satisfied. For my purposes here, charge non-renormalization is crucial, since I wish to interpret these results as a new effect. To verify non-renormalization of charge, I take $\bar k= k$ in $M_{TOT_\mu}$ and use the properties of the background dependent spinors\footnote{For more details see Eq. (\ref{eq:barunkunk}) or the analogous case in\cite{Donoghue:1984zz}.}, to get
\begin{eqnarray}
\bar u_n(\bar k) M_{TOT_\mu} u_n(k)\left|_{\bar k=k} \right.= -e\frac{k_\mu}{E_k}\quad\quad\quad \quad\quad\quad\quad\quad\quad \\ \nonumber +\frac{e}{E_k}\left(\frac{k_\mu}{E_k}\frac{d}{dE_k}-\frac{d }{dk_\mu}\right)\left[\frac{}{}m_eB(k)+k_\mu D^\mu(k)\right]~.
\end{eqnarray}
If I take $\mu=0$, I find
\begin{eqnarray}
\bar u_n(k) M_{TOT_0} u_n(k)\left|_{\bar k=k} \right.=-e~,
\end{eqnarray}
and charge is not renormalized. 

I also verify the Ward Identities, $\Delta k^\mu M_{\mu}=0$, of the expression in Eq. (\ref{eq:MTOT}). Using the fact that $\Delta k_\mu F^\mu(\Delta k)=0$, I find 
\begin{eqnarray}
\Delta k^\mu M_{TOT\mu}=-e\bar u_n(\bar k)\quad\quad\quad\quad\quad\quad\quad\quad\quad\quad\quad\quad\quad \\ \nonumber \times \left[ B(k)-B(\bar k)+\Delta k^\mu \left[\frac{d B(k)}{dk_\mu}+\frac{d B(\bar k)}{d\bar k_\mu}\right] \right.\quad\quad\quad\quad\\+\frac{k_\mu+\bar k_\mu}{2m_e}\left[D_\mu(k)-D_\mu(\bar k)\right]-\frac{\Delta k^\mu }{2m_e}\left[ D_\mu(k)+D_\mu(\bar k)\right] \nonumber   \\ +\left[\left[\frac{d D^\nu(k)}{dk_\mu}+\frac{d D^\nu(\bar k)}{d\bar k_\mu}\right]\Delta k^\mu +D_\nu(k)-D_\nu(\bar k)\right]\qquad \nonumber  \\
\left.\times \frac{\left[\gamma_\alpha,\gamma_\nu\right]_-\Delta k_\alpha}{4m_e} \right] u_n(k)\nonumber ~.
\end{eqnarray}
Because of the background dependent spinors, the application of Dirac's equation is modified by Eq. (\ref{eq:waveFuncLoop}) giving the above expression. The third and fifth rows in the above expression cancel to order $\Delta k^3$.  The fourth row cancels to order $m_{DM}^4$. Since all terms are higher order in $\Delta k$ and $m_{DM}$ than I considered, gauge invariance is shown.

\section{The Hamiltonian}
The last thing I do is calculate the Hamiltonian and then from it I determine the cyclotron and spin frequencies. 

Before proceeding, I first simplify the expression in Eq. (\ref{eq:MTOT}): 
\begin{eqnarray}
iM_{TOT_\mu}=-ie \bar u(\bar k) \left[\gamma_\mu\left[1+\frac{1}{2}\left[\frac{D^0(k)}{E_k}+\frac{D^0(\bar k)}{E_{\bar k}}\right] \right.\right.  \label{eq:MTOTTYPE} \\ \left. \left. -R\frac{m_e}{E_k} \bar I_0(k)-R\frac{m_e}{E_{\bar k}} \bar I_0(\bar k)-\frac{1}{2m_e}\Delta k_\nu \bar I^\nu(k)\right]\right. \nonumber \\
 -\frac{D_\mu(k)}{2m_e}-\frac{D_\mu(\bar k)}{2m_e} +R\left[\bar I_\mu(k) +\bar I_\mu(\bar k)\right ]\nonumber \quad\quad  \\
   +  \left[ I^{\nu}_\mu  -  R  \frac{k_\mu+\bar k_\mu}{m_e} \bar I^\nu(k)  + 2R\slashed{\bar I}(k)\delta^\nu_\mu \right. \nonumber \quad\quad \quad \\ \left.\left. +2m_eR\delta^\nu_\mu I_A(k)\frac{}{}\right] 
 \times \frac{\left[\gamma_\alpha,\gamma_\nu\right]_-\Delta k^\alpha}{4m_e} \right]u(k) ~, \nonumber  \quad\quad 
\end{eqnarray}
with $R=m_{DM}^2/m_e^2$ and 
\begin{eqnarray}
I_A(k)=e^2\chi^2\int \frac{d^4\Pi_q}{(2\pi)^3} \frac{2m_e}{(2q\cdot k)^2} =\frac{\delta m_n}{2m_{DM}^2}\left(\frac{m_e}{E_k}\right)^2~, \\
\bar I_0(k) =e^2\chi^2\int \frac{d^4\Pi_q}{(2\pi)^3} \frac{4E_q m_e^3 }{(2q\cdot k)^3}=\frac{\delta m_nm_e}{2m_{DM}^2} \left(\frac{m_e}{E_k}\right)^3~,
\end{eqnarray}
where the other components of $\bar I_\mu(k)$ are suppressed by the dark matter velocity squared, $\beta_{DM}^2$, and I have only kept the dark matter mass to order $m_{DM}^2$. The other functions in Eq. (\ref{eq:MTOTTYPE}) can also be evaluated, giving
\begin{eqnarray}
&&B(k)=-\frac{1}{2}\delta m_n \left(\frac{m_e}{E_k}\right)^2~,\\
&&D^0(k)= -\delta m_n\frac{m_e}{E_k}~,
\end{eqnarray}
where the other components of $D^\mu(k)$ are also dark matter velocity suppressed and
\begin{eqnarray}
\delta m_n= \frac{e^2\chi^2}{(2\pi)^3}\frac{1}{m_e m_{DM}}\int d^3q \bar n(E_q)~.
\end{eqnarray}

I then apply the correction in Eq. (\ref{eq:MTOTTYPE}) to Dirac's equations, with $A^0=0$ and $\vec{A}=\frac{1}{2}\vec B\times \vec r$, and then use it to find the Hamiltonian. Next, I perform a Foldy-Wouthuysen transformation \cite{Mendlowitz,Donoghue:1984zz}. After this transformation, I keep only the upper component of the spinor. The off-diagonal pieces are dropped, since they are effectively two-loop order. Simplifying the resulting expression, I then get\footnote{For more details about this method, see \cite{Donoghue:1984zz} and Appendix \ref{App:Ham}}  
\begin{eqnarray}
H' 
=E_\beta-\frac{e}{2E_\beta}\left[\vec L \cdot B+\vec\sigma\cdot B\right]\left[1-2R\frac{m_e}{E_k}\bar I^0(k)\right] \label{eq:Ham}\\
+\frac{eR}{2E_p}\left[\frac{|k|^2}{m_e^2}\bar I^0(k)  - 2\bar I^0(k)  -2I_A(k) E_p\right] \vec \sigma\cdot \vec B\nonumber   ~,
\nonumber  
\end{eqnarray}
where I have ignored some corrections proportional to $\hat B \cdot \hat k \hat k\cdot \vec \sigma$, which are relevant when $\vec  k$ is parallel to $\vec B$, and $E_\beta$ includes the thermal corrections to the mass of the electron. 

\section{Results}

It is a little non-trivial to compare to experiment the anomalous magnetic moment of my calculation. This is because both the spin and cyclotron frequencies in a magnetic field are modified. Since the measured quantity is $(\omega_a-\omega_c)/\omega_c$, there will be some cancellation. To circumvent this problem, I directly compare this correction to the experimental errors on the frequencies. 

The cyclotron frequency ( $\omega_c$) and the spin frequency ($\omega_{s\bot}$ for velocities perpendicular to $\vec B$) can effectively be read off from from Eq. (\ref{eq:Ham}) and are
\begin{eqnarray}
\omega_c= \frac{e|B|}{2E_\beta} \left[1-\frac{2Rm_e}{E_p} \bar I^0(k)\right]~, \qquad\qquad\qquad \\
 \omega_{s\bot}= \omega_c\left[1+ \frac{\alpha}{2\pi}\frac{E_p}{m_e}\right. \qquad\qquad \qquad\qquad\qquad \\
+ \left. R\left(\left(2-\frac{|k|^2}{m_e^2}\right)\bar I^0(k)    +2I_A(k) E_p\right)\right]~,\nonumber 
\end{eqnarray}
where I have added the zero background correction to the spin frequency, see \cite{Donoghue:1984zz}. I also assumed the velocity can be taken perpendicular to $\vec B$ to leading order. The quantity which experiments can compare to theory is \cite{Fan:2022eto,Hanneke:2010au}
\begin{eqnarray}
R_f=\frac{\omega_a}{\omega_c}=\frac{\omega_{s\bot}-\omega_c}{\omega_c}\simeq R_{f_0}\left[1+\frac{\delta \omega_a}{\omega_{a_0}}-\frac{\delta \omega_c}{\omega_{c_0}}\right]
\end{eqnarray}
where $\delta \omega_{c,a}$ are the corrections from the background and $R_{f_0}$, $\omega_{a_0}$, and $\omega_{c_0}$ are the SM predicted values. Because $\delta \omega_{a,c}$ are similar in size, the dominant correction comes from $\delta \omega_a$, since $\omega_{a_0}$ is much smaller. 

To compare to experiment, I need to determine $\delta m_n$, which requires an understanding of how to approximate $\bar n(k)$. To approximate this, I start with the dark matter density, $\rho_{DM}$, for a given polarization of the dark photon\footnote{Dark photon dark matter is assumed to be, roughly, an even admixture of all three polarizations since it is massive.}. The dark matter number density, $n_{DM}$, is related to the dark matter density, $\rho_{DM}$, as follows\footnote{The one-third below is because I assume the dark matter is evenly distributed among the three polarizations of the dark photon.}
\begin{eqnarray}
n_{DM}=\frac{1}{3}\frac{\rho_{DM}}{m_{DM}}~,
\end{eqnarray}
This is then related to the occupation number, 
\begin{eqnarray}
\bar n = \frac{1}{3}\frac{ n_{DM}}{\frac{4\pi q^2\Delta q}{(2\pi)^3}}~.
\end{eqnarray}
If I then integrate this over the momentum $\vec q$, I very roughly get 
\begin{eqnarray}
\int d^3q\bar  n(E_q)\simeq \int d^3q  \frac{\rho_{DM}}{m_{DM}}\frac{(2\pi)^3}{12\pi q^2\Delta q} \simeq  \frac{(2\pi)^3\rho_{DM}}{3m_{DM}}~.
\end{eqnarray}
Using this expression, I can then determine $\delta m_n$, 
\begin{eqnarray}
\delta m_n= \frac{4\pi}{3} \alpha \chi^2\frac{\rho_{DM}}{m_e m_{DM}^2}~.
\end{eqnarray}
This is a very rough estimate based on the assumption that $\rho_{DM}$ is constant over a range of velocities equal in width to the dark matter velocity. This approximation can definitely be refined.  However, this will just amount to some order one error on the exclusion limits I will give. Since these limits cover many orders of magnitude, this does not have a significant effect on my conclusions. Thus, I will take this rough approximation and leave a more detailed analysis to later work. 

If I now apply all these simplification, I find 
\begin{eqnarray}
&&\frac{\Delta R_f}{R_{f_0}}\simeq \frac{\delta \omega_a}{\omega_{a_0}}\simeq\frac{(2\pi)^2}{3}\chi^2 \frac{\rho_{DM}}{m_{DM}^2E_k^2}~.
\end{eqnarray}

\begin{figure}[t!]
\begin{center}
\includegraphics[width=.5\textwidth,trim={.77in .1in .77in .1in},clip ]{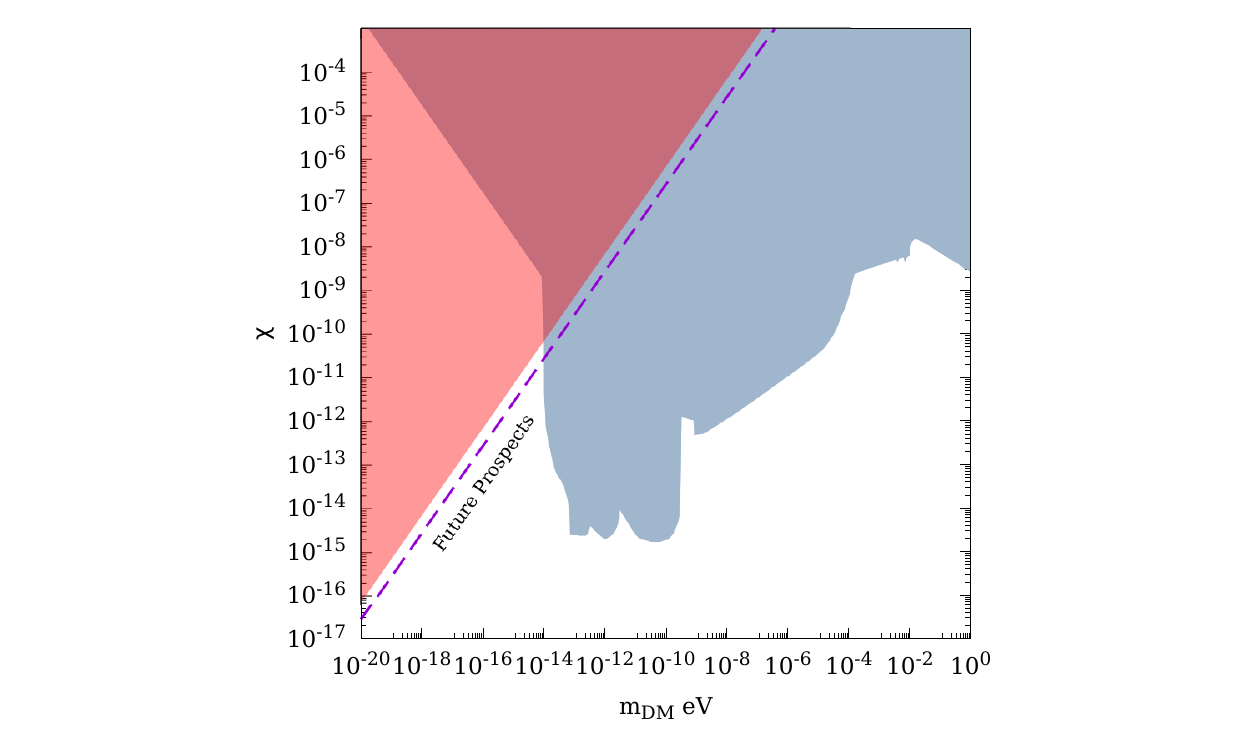}
\end{center}
\caption{The shaded red region is excluded by the fact that it would contribute too much to the anomalous magnetic moment of the electron and the purple dashed line is future prospects of detection. \label{fig:Consstraints}}
\end{figure}
Since the standard model electron anomalous magnetic moment measured in \cite{Fan:2022eto} is consistent with the experimental results, I will assume that $\Delta R_f$ can be no larger than the experimental error on it\footnote{There is a small discrepancy in other measurements of $g-2$. Whether this background effect can explain this discrepancy is left to future work.}. Using the uncertainties on the frequencies found in Figure 4.26 and 4.27 of \cite{Fan:PenningTrap}, 
\begin{eqnarray}
\frac{\Delta \omega_c}{\omega_c}\simeq \pm 2\times 10^{-11}~, \qquad 
\frac{\Delta \omega_a}{\omega_a}=-4\times 10^{-12}~,
\end{eqnarray}
I can then constrain $\Delta R_f$ and in turn put a constraint on $\chi$ for a given $m_{DM}$. Because $\Delta \omega_c/\omega_c$ is the largest error, It will be the dominant contribution to the error of $R_f$ and I get 
\begin{eqnarray}
\frac{\Delta R_f}{R_{f_0}}\simeq \frac{2\pi}{3}\alpha\chi^2 \frac{\rho_{DM}}{m_{DM}^2E_km_e}<\frac{\Delta \omega_c}{\omega_c} = 4\times 10^{-11}~.
\end{eqnarray}
From this, I find that unless the gauge mixing parameter satisfies the following relation:
\begin{eqnarray}
\chi< 7.1\times 10^{3} \frac{m_{DM}}{eV}~,
\end{eqnarray}
it is excluded, where I have used $\rho_{DM}\simeq 0.3$ GeV/cm$^3$, and $|\vec k|\simeq 7.2$ MeV, see \cite{Fan:2022eto}.

In Fig. (\ref{fig:Consstraints}), I give the new constraints on dark photon dark matter in red, the future prospects are a dashed purple line, and the existing constraints are in gray, which is from \cite{Caputo:2021eaa}. 

This same effect will appear for ultralight axion-like particles. For axion-like particles with a $\gamma_5$ coupling, the function $C(k)$ and $D^{\mu}(k)$ are one-half those found in Eq. (\ref{eq:Cofk}) and (\ref{eq:Dofk}) while $B(k)=0$. This means the contribution to the anomalous magnetic moment parameterized by $D^\mu(k)$ will be just one-half that found here. The contribution coming from the corresponding $F_\mu(\Delta k)$ is also of the same form. Thus, I expect a similar type of constraint. The details of this calculation and some applied phenomenology will be discussed in future work. However, this discussion has emphasized the need to consider the Bose enhancement of all kinds of ultralight dark matter backgrounds.  

\section{Conclusions}
When dark matter is ultralight, its number density is extremely high and zero temperature field theory is no longer viable. Since ultralight dark matter is bosonic in nature, the number of particles which can fill a given state is unbounded.  If the occupation number of the background states is large, the Bose enhancement effect can be enormous. This effect is not only important for decay processes, as is well known, but also any loop diagram, as shown here. In this letter, I examined the effect of background dark matter on the electron anomalous magnetic moment for an ultralight dark photon. Because the anomalous magnetic moment of the electron is so precisely measured and the Bose enhancement so large for ultralight dark matter, significant constraints were placed on the dark photons gauge mixing for a give mass. As briefly mention, this same procedure can be applied to other ultralight dark matter which interacts with the electron like axion-like particles. 

\section*{Acknowledgement}
I would like to thank Ariel Arza, Tsutomu T. Yanagida, Kim Siang Khaw,  and Keith A. Olive for useful discussions in regards to this work. J. L. E. is supported by a start-up grant from Shanghai Jiao-Tong University

\appendix 

\section{The Vertex Correction\label{App:Loop}} 
Using the parameterization of the self-energy in Eq. (\ref{eq:waveFuncLoop}) allows us to define the inverse of the corrected fermion propagator as follows\footnote{Here, $m_e$ will signify the renormalized mass coming from zero temperature field theory.}
\begin{eqnarray}
S^{-1}(k)=\slashed{\tilde k} -\tilde m=\slashed{k}-m_e+\Sigma_n(k) \qquad\qquad\qquad \\=\left[1+C(k)\right] 
\times\left[\left[1+\frac{1}{E_k}D^0(k)\right]E_k\gamma_0 \nonumber \right. \\ \left.-\left[1+\frac{\vec k \cdot \vec D(k)}{\vec k^2}\right]\vec\gamma\cdot \vec k \frac{}{}-\left[m-B(k)\right] \right]~,\nonumber \label{eq:propDefpmu}
\end{eqnarray}
where the $1+C(k)$ is the wave function renormalization and so has been factored out. When the background contribution to the self-energy is included, the renormalized propagator can then be defined, in terms of the implicitly defined $\tilde k_\mu$ above, as 
\begin{eqnarray}
S^R(x-y)=i\int \frac{d^4k}{(2\pi)^4}\frac{Z_2^{-1}(\slashed{\tilde k}+\tilde m_e) e^{-ik\cdot (x-y)}}{\tilde k^2-\tilde m^2+i\epsilon}~.\label{eq:waveProp}
\end{eqnarray}

To get $Z_2^{-1}$, I compare the above expression to what I would get if I calculated $\langle \psi_n(x)\bar \psi_n(y)\rangle$. However, before I can do this, I must define my background dependent spinors, $u_n(k)$. Without taking background dependent spinors, charge non-renormalization and gauge invariances would appear to be violated, as discussed in \cite{Donoghue:1984zz}. There, they took temperature dependent spinors for the same reason. 
The background dependent spinors are chosen to satisfy the following modified Dirac equation,
\begin{eqnarray}
\left[\slashed{k}-m_e+\Sigma_n\right]u_n(k)=0~,\label{eq:EOMunSpinor}
\end{eqnarray}
and have the following properties
\begin{eqnarray}
\bar u_n(k)\gamma_\mu u_n(k) = \frac{\tilde k_\mu}{\tilde E_k}~,\quad\quad\quad \bar u_n(k) u_n(k)=\frac{\tilde m}{\tilde E_k}~,\label{eq:barunkunk}
\end{eqnarray}
and
\begin{eqnarray}
\sum\limits_{\rm spin} u_n(k)\bar u_n(k) = \frac{\slashed{\tilde k}+m_e}{2\tilde E}~,
\end{eqnarray}
where
\begin{eqnarray}
\tilde E =\left(1+\frac{D^0(k)}{E_k}\right)~,\label{eq:Etil}
\end{eqnarray}
is the background corrected energy\footnote{The wave function renormalization piece is removed from Eq. (\ref{eq:Etil}), since it cancels from all calculations}.

With these background dependent spinor properties, I can define the free fermion field $\psi(x)$ in a background and calculate $\langle  \psi(x)\bar \psi(y)\rangle$ . Comparing this propagator to that in Eq. (\ref{eq:waveProp}), I can determine the wave function renormalization, which is\footnote{This is found using techniques analogous to those used in \cite{Donoghue:1984zz}} 
\begin{eqnarray}
Z_2^{-1}\!= 1+C(k)+\frac{m_e}{E}\frac{d}{dE}\!\left[B(k)+\frac{k_\mu D^\mu(k)}{m_e}\right]\!-\frac{D^0(k)}{E} ~.
\end{eqnarray}

Now, I calculate the three diagrams in Fig. (\ref{fig:VerCorrections}). First, I calculate the self-energy vertex correction\footnote{There are some subtleties which lead to the derivatives of $B(k)$ and $D^{\mu}(k)$, which relate to the definition of the energy in the background field. A similar procedure was used in \cite{Donoghue:1984zz} when calculating in a thermal background.}

\begin{eqnarray}
iM_{SE_\mu}=\qquad\qquad\qquad\qquad\qquad\qquad\qquad\qquad\qquad \\ ie\bar u_n(\bar k)\left[ C(\bar k) +\frac{1}{E_{\bar k}}\left[m_e\left.\frac{\partial B(\bar k)}{\partial E_{\bar k}}\right|+\left.\frac{\partial \bar k_\mu D^\mu(k)}{\partial E_{\bar k}}\right|\right]\right.\nonumber \\ \nonumber 
\left.-\frac{D^0(\bar k)}{E_k} +\left[\left.B(\bar k)\right|  + \left.\slashed{D}(\bar k)\right| \right] \frac{1}{\slashed{\bar k}-m_e}\right]\gamma_\mu u_n(k)~,
\end{eqnarray}
where a $\left. \right|$ indicates $\bar k^2=m_e^2$ here and throughout the draft.

Next, I calculate the background dependent mass counterterm vertex correction. This contribution is essential since I am considering background dependent spinors and is
 \begin{eqnarray}
iM_{CT_\mu}=\!-ie\bar u_n(\bar k) \left[\left.B(\bar k)  + \slashed{D}(\bar k)\right]\right|\frac{1}{\slashed{\bar k}-m_e}\gamma_\mu u_n(k).
\end{eqnarray}

The last thing I need to calculate is the one-loop correction to the vertex. Instead of just calculating the vertex diagram, I will rely on the fact that if $k=\bar k$, then I have  
\begin{eqnarray}
iM_{VER}\left|_{\bar k =k}\right. = -ie\bar u_n(k)\frac{d\Sigma_n}{dk_\mu}u_n(k)~.
\end{eqnarray}
Using this fact, I can decompose the one-loop correction into a sum of derivatives of the self-energy corrections with respect to $k$ and $\bar k$ and the leading order correction in $\Delta k=\bar k-k$, as follows
\begin{eqnarray}
iM_{VER} = \qquad\qquad\qquad\qquad\qquad\qquad\qquad\qquad \\ \nonumber -i\frac{e}{2}\bar u_n(\bar k)\left[\frac{d\Sigma_n(k)}{dk^\mu} + \frac{d\Sigma_n(\bar k)}{d\bar k^\mu}+F_\mu (\Delta k)\right]u_n(\bar k)~.\label{eq:VerCorr}
\end{eqnarray}
where\footnote{Below I have neglected the derivative of $C(k)$ since it will not contribute at the one-loop level when $\slashed{k}-m_e$ is projected on $u_n(k)$.}
\begin{eqnarray}
\frac{d\Sigma_n(k)}{dk_\mu}=\gamma_\mu C(k)+\gamma_\nu \frac{d D^\nu(k)}{dk_\mu}+\frac{dB(k)}{dk_\mu}~,\label{eq:DiffSig}
\end{eqnarray}
and
\begin{eqnarray}
\!\!\!\!\! F_\mu(\Delta k)\quad\quad\quad\quad\quad\quad\quad\quad\quad\quad\quad\quad\quad\quad\quad\quad\quad\quad\quad \\ \nonumber 
=  -e^2\chi^2\int \frac{d^4\Pi_q}{(2\pi)^3}\frac{2\left(\slashed{q}+m_e\right)\left[\Delta \slashed{k},\gamma_\mu\right]-4\gamma_\mu\left[\Delta k_\mu q^\mu\right]}{\left(q^2+2 q_\mu k^\mu\right)^2} ~. \label{eq:FDelK}
\end{eqnarray}

I can now use Gordon Decomposition on the $\gamma_\nu$ in Eq. (\ref{eq:VerCorr}), ignore terms of order $\Delta k^2$, and simplify to get\footnote{The following expression reproduces the standard zero temperature calculation with a photon if I define $B(k)$, $D^\mu(k)$, and $F(k,\Delta k)$ in terms of the zero temperature photon propagator and take $\chi=1$} 
 \begin{eqnarray}
iM_{VER}= -\frac{ie}{2}\bar u( \bar k) \left[\left(C(k)+C(\bar k)\right)\gamma_\mu\right.\qquad \qquad \qquad \\   \nonumber  +\frac{d B(k)}{dk_\mu}+\frac{d B(\bar k)}{d\bar k_\mu} +\frac{k_\nu}{m_e}\frac{d D^\nu(k)}{dk_\mu} +\frac{\bar k_\nu}{m_e}\frac{d D^\nu(\bar k)}{d\bar k_\mu} \\  
\left. +\frac{\left[\gamma_\alpha,\gamma_\nu\right]_-\Delta k_\alpha}{4m_e}\left[\frac{d D^\nu(k)}{dk_\mu} +\frac{d D^\nu(\bar k)}{d\bar k_\mu}\right]+F(\Delta k)\right]u(k) ~.\nonumber 
\end{eqnarray}

The sum of these diagrams then gives the expression found in Eq. (\ref{eq:MTOT}). It is sometimes useful to not Gordon decompose the $\gamma_\nu$ in Eq. (\ref{eq:DiffSig}). In this case the Ward identities are much more trivial to verify and can be shown to be true to order $\Delta k^3$, higher order than I consider, and all order in $m_{DM}$.

\section{The Hamiltonian \label{App:Ham}}

Next, I examine the equation of motion for the electron including the corrections found in Eq. (\ref{eq:MTOTTYPE}),  
\begin{eqnarray}
\left[\frac{}{}\slashed{k}+\gamma_0 D^0 -m_e+B(k) \right.\qquad\qquad\qquad\qquad\qquad \qquad    \\ \nonumber -e\vec\gamma \cdot \vec A\left[1+\frac{D^0(k)}{E_k}-2R\frac{m_e}{E_k}\bar I^0(k)\right] \qquad \quad\,\, \\  - e\left(\Delta k^j A^i\right)\left[ 2 R  \frac{k^i}{m_e} \bar I^0(k) \delta^\nu_0 + 2R\gamma_0 I^0(k)\delta^\nu_i \right.\nonumber  \\ \left.\left. \frac{}{}+2m_eR\delta^\nu_i I_A(k)\right] \frac{\left[\gamma_j,\gamma_\nu\right]_-}{4m_e}
\right]u_n(k)~, \qquad\qquad \,\,\,\,   \nonumber 
\end{eqnarray}
where I used the fact that $A_0=0$ and $A=\frac{1}{2} \vec{r}\times \vec B$. I now solve for the energy $k^0$, which is effectively the Hamiltonian, to get
\begin{eqnarray}
H=\left[\frac{}{}-\vec\sigma\cdot \vec \pi_\beta\rho_1- D^0 +\rho_3\bar m_e\right. \qquad\qquad\qquad\qquad\qquad  \\ + \frac{eR}{m_e}\left[    \bar I^0(k)\frac{1}{2m_e}\left(\vec\sigma\times \vec B\right)\cdot \vec k \rho_2\right.\qquad\qquad \nonumber \\
\left. \left. - \left[I^0(k)+m_eR I_A(k)\rho_3\right]\vec\sigma\cdot \vec B 
\right] \right]~,\qquad \nonumber
\end{eqnarray}
where
\begin{eqnarray}
\bar m_e= m_e-B(k)~,
\end{eqnarray}
and \footnote{The matrices below are defined in $2\times 2$ dimensional product space with the Pauli matrices}
\begin{eqnarray}
\rho_1=-\gamma_5 \quad\quad \rho_2= i\gamma_0\gamma_5 \quad\quad \rho_3=\gamma_0~.
\end{eqnarray}

To determine the correction to the frequencies, I first make a Foldy-Wouthuysen transformation to the Hamiltonian, as was done in \cite{Mendlowitz,Donoghue:1984zz}, 
\begin{eqnarray}
H'= \exp\left[-i\frac{1}{2} \phi \rho_2\right]H\exp\left[i\frac{1}{2} \phi \rho_2\right]~,
\end{eqnarray}
and take
\begin{eqnarray}
\tan\phi=\frac{\vec\sigma \cdot \pi_\beta}{\bar m_e}~.
\end{eqnarray}
Simplifying, I get 
\begin{eqnarray}
H' 
\simeq E_\beta-\frac{e}{2E_\beta}\left[\vec L \cdot B+\vec\sigma\cdot B\right]\left[1-2R\frac{m_e}{E_k}\bar I^0(k)\right]\\
+\frac{eR}{2E_p}\left[\frac{|k|^2}{m_e^2}\bar I^0(k)  - 2\bar I^0(k)  -2I_A(k) E_p\right] \vec \sigma\cdot \vec B\nonumber \\
+\frac{eR}{2E_p}\left[-\frac{|k|^2}{m_e^2}\bar I^0(k)-2\bar I^0(k)\left(\frac{E_p}{m_e}-1\right)  \nonumber \right. \\ \left. \frac{}{}- 2I_A(k) \left(E_p-m_e\right)\right]\vec\sigma\cdot \hat k \hat k\cdot \vec B   ~. 
\nonumber  
\end{eqnarray}

\end{document}